\documentclass[%
 reprint,
 amsmath,amssymb,
 aps,
 prab,
]{revtex4-2}
\usepackage[utf8]{inputenc}
\usepackage[T1]{fontenc}
\usepackage{mathptmx} %
\usepackage{amssymb}
\usepackage{booktabs}
\usepackage{bm}

\usepackage[pdftex]{graphicx} %
\usepackage[pdftex,linkcolor=black,pdfborder={0 0 0}]{hyperref} %
\usepackage{calc} %
\usepackage{enumitem} %

\usepackage[all]{nowidow} %
\usepackage[protrusion=true,expansion=true]{microtype} %

\usepackage{lipsum} %
\usepackage{acronym}
\usepackage{xcolor}
\usepackage{siunitx}
\usepackage{verbatim}
\usepackage{physics}
\usepackage[noabbrev, capitalise]{cleveref}

\usepackage{layouts}

\newacro{RL}{reinforcement learning}
\newacro{ML}{machine learning}
\newacro{FPGA}{field programmable gate array}
\newacro{NN}{neural network}
\newacro{BPM}{beam position monitor}
\newacro{FIR}{finite impulse response}
\newacro{RF}{radio frequency}
\newacro{AIE}{AI engine}
\newacro{ReLU}{rectified linear unit}
\newacro{HBO}{horizontal betatron oscillation}
\newacro{PPO}{proximal policy optimization}
\newacro{PCG}{permuted congruential generator}
\newacro{KARA}{Karlsruhe research accelerator}
\newacro{THD+N}{total harmonic distortion plus noise}
\newacro{HLS}{high level synthesis}
\newacro{DAC}{digital to analog converter}
\newacro{DMA}{direct memory access}
\newacro{MDP}{Markov decision process}
\newacro{SAC}{soft actor-critic}
\newacro{BBB}{bunch-by-bunch}
\newacro{POMDP}{partially observable Markov decision process}
\newacro{AI}{artificial intelligence}
\newacro{IP}{intellectual property}

\hypersetup{ 	
pdfsubject = {},
pdftitle = {},
pdfauthor = {}
}

\begin{document}
\title{Microsecond-Latency Feedback at a Particle Accelerator by Online Reinforcement Learning on Hardware}
\author{Luca~Scomparin}
  \email{luca.scomparin@kit.edu}
\author{Michele~Caselle}
\author{Andrea~{Santamaria Garcia}}
\author{Chenran~Xu}
\author{Edmund~Blomley}
\author{Timo~Dritschler}
\author{Akira~Mochihashi}
\author{Marcel~Schuh}
\author{Johannes~L.~Steinmann}
\author{Erik~Bründermann}
\author{Andreas~Kopmann}
\author{Jürgen~Becker}
\author{Anke-Susanne~Müller}
\author{Marc~Weber}
\affiliation{Karlsruher Institut für Technologie, Kaiserstraße 12, 76131 Karlsruhe, Deutschland}
\date{\today}

\begin{abstract}

    The commissioning and operation of future large-scale scientific experiments will challenge current tuning and control methods. \Ac{RL} algorithms are a promising solution thanks to their capability of autonomously tackling a control problem based on a task parameterized by a reward function. The conventionally utilized \ac{ML} libraries are not intended for microsecond latency applications, as they mostly optimize for throughput performance. On the other hand, most of the programmable logic implementations are meant for computation acceleration, not being intended to work in a real-time environment. %
    To overcome these limitations of current implementations, \ac{RL} needs to be deployed on-the-edge, i.e. on to the device gathering the training data. In this paper we present the design and deployment of an experience accumulator system in a particle accelerator. In this system deep-\ac{RL} algorithms run using hardware acceleration and act within a few microseconds, enabling the use of \ac{RL} for control of ultra-fast phenomena. The training is performed offline to reduce the number of operations carried out on the acceleration hardware. The proposed architecture was tested in real experimental conditions at the \ac{KARA}, serving also as a synchrotron light source, where the system was used to control induced horizontal betatron oscillations in real-time. The results showed a performance comparable to the commercial feedback system available at the accelerator, proving the viability and potential of this approach. %
    Due to the self-learning and reconfiguration capability of this implementation, its seamless application to other control problems is possible. Applications range from particle accelerators to large-scale research and industrial facilities.

    \acresetall %
\end{abstract}

\maketitle

\section{Introduction}
The tuning of future large-scale scientific experiments will pose great challenges. Due to the sheer amount of parameters that require adjustment, manual tuning will be an extremely demanding task. This will in turn impact the budget of large scientific endeavors as the resource requirements for commissioning and operation of the facility will potentially become unbearable. Automatic algorithms capable of finding an optimum in a parameter space, also known as optimizers, are a possible solution and have already been successfully applied to particle accelerators~\cite{mustapha2009optimization,Scheinker2013,Hofler2013}, with some attempts of creating standardized interfaces and algorithm libraries~\cite{xopt}. Even for these more refined techniques, the increase in the number of variables involved in the task will reduce their performance, a phenomenon known as the \emph{curse of dimensionality}. 

One of the possible solutions is the use of data-driven \ac{ML} techniques~\cite{Edelen2020,Roussel2021,Kaiser2024}. In the case of control problems, a promising approach is the use of \ac{RL}, a paradigm where an agent is trained to learn a policy in order to maximize a reward function, encoding a control goal, acting on an environment based on a set of observations. \ac{RL} algorithms have already been proven to be a powerful tool through their application to control problems in several large scale facilities~\cite{Degrave2022,learningtodovswhiledoing,madysa2022automated,bruchon2020basic,PhysRevAccelBeams.23.124801,o2020policy,PhysRevAccelBeams.24.104601}.

Typically, a controller must operate within time scales comparable to those of the controlled dynamics. The fact that the evaluation of a control policy needs to be performed in a scheduled time frame is by definition a real-time constraint~\cite{real_time_survey}. Failure to meet this kind of requirement means the controller could be partially effective or even completely ineffective. As such, when the inference time reaches the microsecond timescale, the use of conventional CPUs becomes challenging, as the several layers of abstraction and caching between the application and hardware limit the achievable latency performance. Specifically, the latency achieved can not only be higher than the requirements, but its high variance could produce spurious, unpredictable violations of these constraints. Different kinds of computing hardware can circumvent these limitations by combining a higher degree of parallelization with a system of lower intrinsic overhead. Modern heterogeneous computing platforms are a good candidate for this, combining conventional CPUs with a \ac{FPGA} and an \ac{AI} accelerator. An \ac{FPGA} allows the definition of custom programmable digital logic that can reach levels of parallelism that are constrained only by the number of programmable cells available on the device. Moreover, they allow precise control over the timing of each operation. The \ac{AI} accelerator allows the efficient execution of more specific tasks (e.g. floating point vector multiplication), for which \acp{FPGA} are less suitable. These three components share memory, allowing a customized and optimized data-flow architecture.

One of the main issues that prevent the application of \ac{RL} to many large-scale facilities is their sample efficiency, the amount of interaction with an environment required to fully train an agent. This means that the data necessary to train a successful agent is usually difficult to obtain in the real world due to their low repetition rates~\cite{Hirlaender:2023ivj}. This issue can be mitigated by pre-training on a simulation, with the risk that the agent might perform sub-optimally in case the simulation differs significantly from the real world~\cite{learningtodovswhiledoing}. A different approach is to use more sample efficient algorithms, but those tend to complicate the tuning of the hyper-parameters~\cite{PhysRevAccelBeams.23.124801}. The benefit of bringing \ac{RL} to facilities operating on microsecond timescales, like numerous synchrotron light sources, MHz-rate free electron lasers and circular colliders, is that sufficient training data can be generated in real time, bypassing the need of training in simulation. Conversely, the constraints on the time taken to select an action require a more careful implementation of the agent.

In this work, we developed a closed-loop \ac{RL} feedback system that was successfully deployed and commissioned at the \ac{KARA}, an accelerator test facility and synchrotron light source. 
This novel scheme enables direct \ac{RL} online training in the microsecond time domain, setting the precedent as the first implementation of its kind in particle accelerators.

\section{Real-time reinforcement learning}
In \ac{RL} the environment is modeled as a \ac{MDP}, a 4-tuple $(S, A, P, R)$ where $S$ is the state space, $A$ is the action space, $P$ are the transition dynamics, and $R$ is the set of all rewards.
The environment provides a scalar reward value $R$ that encodes the goal of the control problem. The goal of an \ac{RL} agent is to maximize the expected return $G$, defined as the cumulative reward in all future steps

\begin{equation}
G_t = \sum_{k=0}^{\infty} \gamma^k R_{t+k+1},
\end{equation}
where $\gamma \in [0,1]$ is the discount factor, used to bound the cumulative reward in infinite horizon problems, extending the \ac{MDP} to $(S, A, P, R, \gamma)$.

This framework models decision-making in partially stochastic control processes and incorporates the Markov property, where the conditional probability distribution $P$ of future states depends only on the present state.

Often the full state of the environment is not directly accessible and needs to be constructed from observations, giving place to \ac{POMDP} defined as $(S, A, P, R, \gamma, O, \epsilon)$, where $O$ is the space of possible observations and $\epsilon$ the probability of transitioning to a new state $s_t$ when $o_t$ is observed $\epsilon = \mathbb P(s_t|o_t)$.

Many modern \ac{RL} algorithms are based on an actor-critic architecture, where an actor function, responsible for choosing the action, is paired with a critic function, estimating the expected return of a given state. The critic can either implement a \emph{value function} $v_\pi: S \rightarrow \mathbb{R}$ or \emph{action-value function} $Q_\pi: S \times A \rightarrow \mathbb{R}$, depending on whether it also takes into account the action that is going to be taken or not~\cite{10.5555/3312046}.  In modern deep-\ac{RL} both actor and critic are approximated with a \ac{NN}.

It is worth noticing that, given the interaction with an active environment, most non-simulated \ac{RL} agents are subject to real-time constraints. Depending on the time scale of the environment, this requirement could be addressed by simply allocating enough computing power to the agent. When the latency requirements become more demanding, the compound overhead of transferring data and evaluating the agent can mandate more adaptable computing architectures like \acp{FPGA} or heterogeneous platforms, in order to better control the scheduling of the data-processing pipeline.

\subsection{Related work: RL on FPGA}
A brief outline of the currently available implementations of \ac{RL} on \ac{FPGA} is described below. A more comprehensive review can be found in reference \cite{rlfpgasurvery}. There are two main classes of implementations, depending on the nature of the value function and policy. Several works store the action-value function in a tabular form, and based on this choose the action with the maximum expected cumulative reward for each state \cite{9211770, 10.1145/3453688.3461533, 8574886, 9150141, 8937555}. The main issue of this kind of system is that its resource usage grows exponentially with the size of the observation and action space, while being not directly applicable to environments with continuous action and observation spaces. These kind of environments are widely used in particle accelerator applications and better fit the problem of this work.

A different approach is deep-\ac{RL}, where the value function and policies are approximated with a \ac{NN} \cite{8445099, 10035176, fa3crl, 10.1145/3039902.3039915, 9460695, 10.1145/3322645.3322693, 9114846}. This approach scales better with increasing size of the observation and action vector, but tends to have a more complex training routine. These implementations lack the flexibility of choosing different training algorithms and changing the \ac{NN} topology during operation. Distinct algorithms and their respective hyperparameter selections might exhibit different performances based on the problem at hand. Therefore, the capability of flexibly choosing these components greatly reduces the deployment effort and expedites the development process. 
Moreover, the policy \ac{NN} implementations shown in the references \cite{9211770, 10.1145/3453688.3461533, 8574886, 9150141, 8937555,8445099, 10035176, fa3crl, 10.1145/3039902.3039915, 9460695, 10.1145/3322645.3322693, 9114846} do not have real-time applications in mind, and as such the application to real-time particle accelerator controls would be limited.

For these reasons, we have chosen to implement a deep-\ac{RL} system capable of performing such dynamic reconfiguration and naturally fulfilling real-time constraints.

\subsection{Experience accumulator architecture}
\begin{figure}[b]
    \centering
    \includegraphics[width=\linewidth]{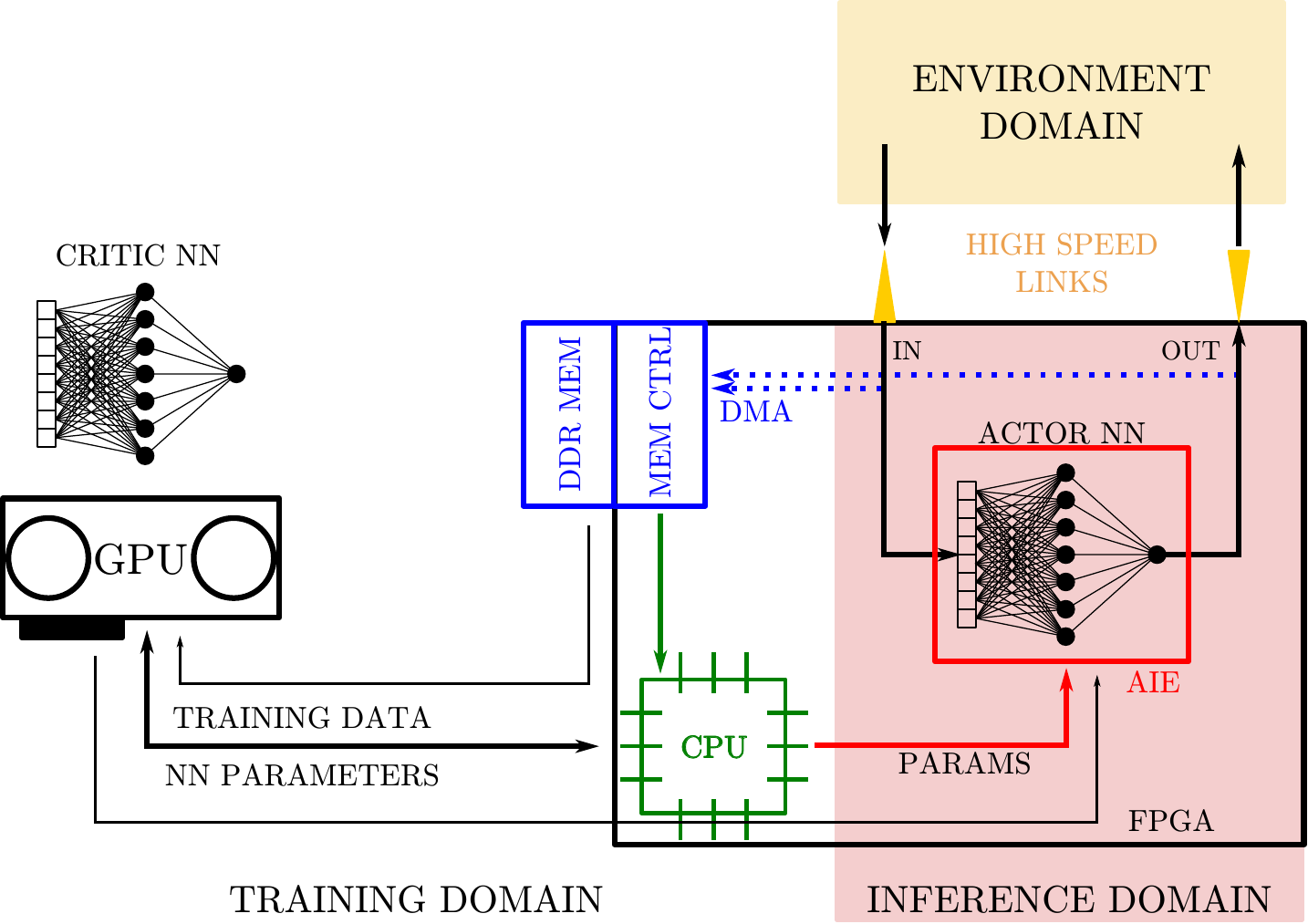}
    \caption{Schematic drawing of the experience accumulator architecture showing the partition between real-time or inference domain (on the right) and training-time domain (on the left) on a heterogeneous platform.}
    \label{fig:experience_accumulator}
\end{figure}

In this study, the system is implemented by employing an \emph{experience accumulator} architecture. Depending on the complexity of the algorithm, the \textit{training} processes in modern deep-\ac{RL} can take a large amount of time compared to the time required for inference of the policy. As such, it is not possible to perform them at a microsecond level timescale. A possible solution is to only implement a general real-time policy $\pi^{(\text{edge})}$ for \textit{inference}, using cutting-edge low-latency computing platforms. 
During application time, the real-time policy network only needs to perform \textit{forward passes}, i.e. predicting the next action $a_i$ based on the observed signal $o_i$.
The interactions of the policy \ac{NN} with the environment are recorded, providing state-action-reward tuples $\{(s_i, a_i, r_i)_{i=1,\dots,T}\}$, that can be used for asynchronously training an emulated copy of the agent $\pi^{(\text{CPU})} \gets \pi^{(\text{edge})}$ on conventional computing platforms, such as a CPU. As such it is possible to achieve a greater degree of flexibility and abstraction (\Cref{fig:experience_accumulator}). This comes at the expense of higher overhead during training, while still maintaining low inference latency. For the \ac{PPO} algorithm agent used in this work, only the actor \ac{NN} is implemented on hardware, while the critic can be evaluated at training time.

A similar approach is presented in a previous work~\cite{exp_acc_old}, despite applying it to a system with less stringent latency requirement, thus not requiring heterogeneous computing platforms but an off-the-shelf x86 processor. Additionally, in this study, we further extend the capability of the experience accumulator architecture by introducing a post-hoc reward definition. In the definition of \ac{RL} control problems, the reward is provided on a step-by-step basis by the environment. In the case of a high-speed feedback system, interacting with a real-time environment, it would then become necessary to perform this computation at pace with the environment. The need for a real-time reward function implementation increases the complexity of the system, as it needs to be implemented in the \ac{FPGA} or AI accelerator, requiring specialized expertise.

In the experience accumulator architecture, a further simplification is possible. The rewards obtained by an agent are only necessary during its training. If the reward is a function only of the observation and action vectors, it is possible to skip its real-time computation and perform it at training time based on the stored observation-action tuples. This \textit{Training Time Reward Definition} technique reduces development efforts as it allows the function to be implemented on a CPU, where the expertise required is more commonly available. In the implementation used in this work, the reward function is developed in Python and specified by the experimenter before starting the training, in this way simplifying the \emph{reward engineering}.

\section{Closing the RL Loop: The KINGFISHER system}
In order to simplify the deployment of \ac{RL} policies in an experience accumulator architecture, all the \ac{FPGA} infrastructure necessary for interfacing with the particle accelerator and storing the agent experience were implemented into an AMD-Xilinx Extensible Platform called KINGFISHER~\cite{Scomparin:2022clg}. This approach has the effect of decoupling the low-level operations allowing the development of the policy and feature extraction algorithm in high-level programming languages, thanks to the aid of tools like \ac{HLS}.

The device of choice is the AMD-Xilinx Versal VCK190~\cite{vck190}, a novel heterogeneous computing platform comprising a \ac{FPGA} and ARM processor in a system-on-chip architecture, with an \ac{AIE} array capable of accelerating multiplication intensive tasks, as, for instance, \ac{NN} inference.

The KINGFISHER platform receives data via a $\SI{40}{Gbps}$ Aurora 64b/66b link~\cite{aurora}. This data can be used to create the observation data-stream. The action data-stream is used to control a \ac{DAC} to produce an analog control signal. For the purpose of future experiments special digital output interfaces are also available. The observation-action data-streams are written to the DDR memory, without active intervention of the CPU, by means of a \ac{DMA} block implemented in the \ac{FPGA}. The ARM processor in the Versal core then copies the data in the DDR to a file that is made available over the network. Moreover, a control server makes the parameters of the system available via the EPICS~\cite{epics} control system. A trigger input is available to start the agent action at a precise time. Moreover, a counter allows the interaction with the environment only for a limited number of steps.

In order to guide the exploration of \ac{RL} algorithms, a stochastic component is usually needed. Thus an \ac{IP} core implementing the \ac{PCG} algorithm~\cite{oneill:pcg2014} produces a continuous stream of 32-bit floating pseudorandom-numbers uniformly distributed between $0$ and $1$ at a rate of $\SI{125}{MSps}$.

\begin{figure*}[htb!]
    \centering
    \includegraphics[width=0.8\linewidth]{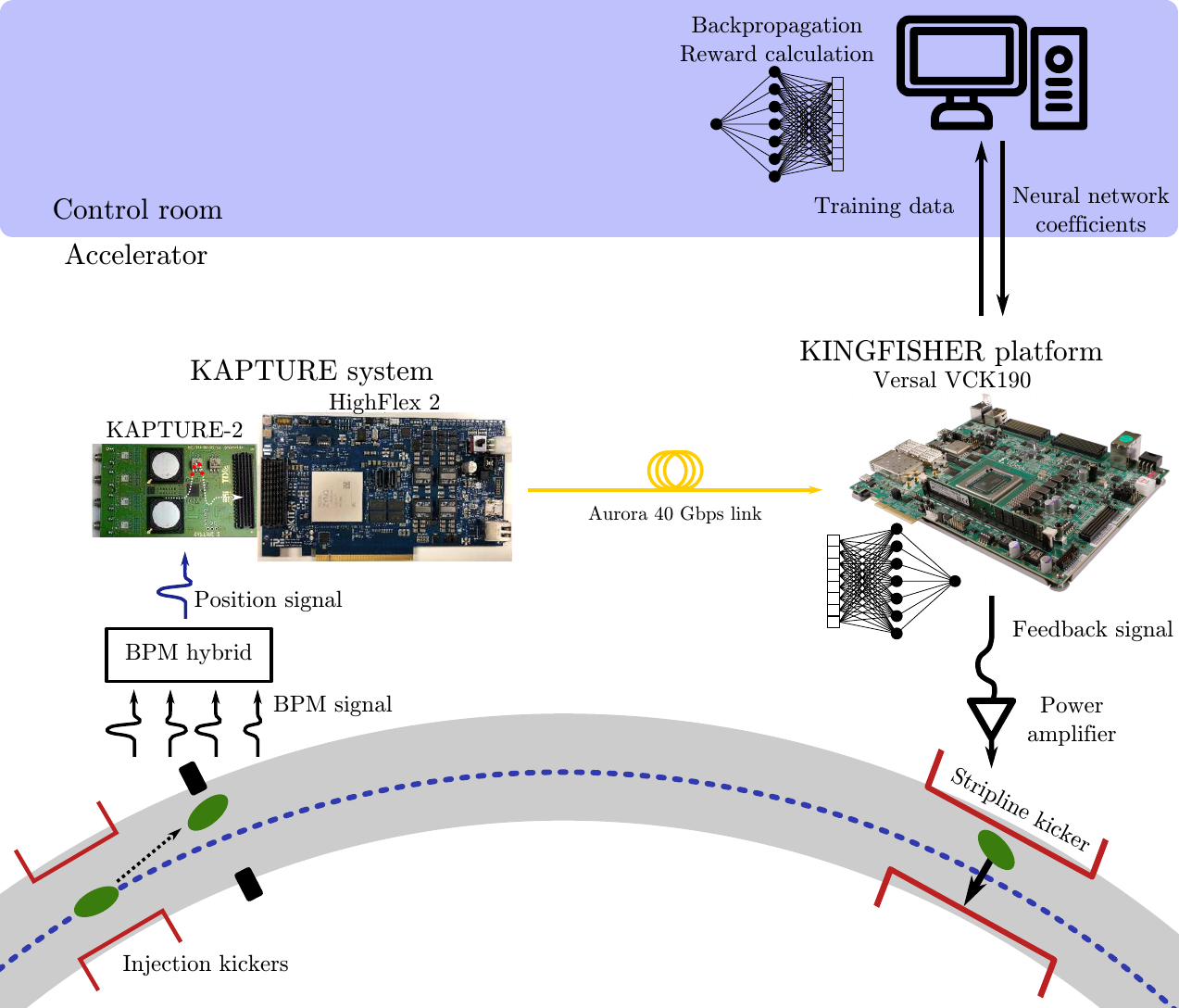}
    \caption{Drawing of the hardware infrastructure employed in this work. The bunch position in the beam pipe is measured by processing a \ac{BPM} signal with an analog hybrid. The produced analog signal is sampled with KAPTURE and then forwarded by an HighFlex 2 board through high-speed links to a Versal device. The KINGFISHER system programmed on this Versal then connects this data stream into the RL controller, while applying the output action to a stripline kicker. Every episode, comprised of $2048$ interaction steps with accelerator, data is sent back to the control room for training.}
    \label{fig:top_schematic}
\end{figure*}

For the control of the \ac{HBO} the specific deployment configuration is shown in \cref{fig:top_schematic}. To sample the fast-peaked signals with high MHz rates commonly generated by diagnostic instrumentation at accelerator-based light sources like synchrotrons, the KIT-developed KAPTURE system~\cite{kapture} is utilized. This system functions as a digitizer, enabling the sampling of four channels at a frequency equal to the accelerator's \ac{RF}, with a selectable sampling interval down to $\SI{3}{ps}$. The system utilizes an Highflex card with the capability of performing transfers directly to a GPU with the GPUDirect technology~\cite{Caselle_2017}. This setup provides bunch-by-bunch and turn-by-turn data. In order to obtain the bunch position relevant for the control of the \ac{HBO}, this infrastructure is used to sample the horizontal position signal from an analog Dimtel BPMH-20-2G hybrid~\cite{bpmhybrid} combining the four signals from a button \ac{BPM}. The action analog signal is fed into a broadband amplifier and then applied to a stripline kicker used to influence the beam. 

\subsection{Implementing Reinforcement Learning Algorithms on KINGFISHER}
The algorithm used in the current work is \ac{PPO}~\cite{schulman2017proximal}. This choice was dictated by its stability with respect to the change of hyperparameters. The reduced sample efficiency compared to off-policy algorithms like \ac{SAC} does not affect the current application, as it is counterbalanced by the high experience collection rate. Other \ac{RL} algorithms are nonetheless easy to integrate thanks to the experience accumulator architecture.

The actor network is implemented in the \ac{AIE}. The employed \ac{PPO} algorithm implementation uses a \ac{NN} to select the mean value of a Gaussian distribution, from which the action applied to the environment is chosen. The standard deviation of the probability distribution is a trainable parameter, that is updated together with the \ac{NN} coefficients. 

\begin{figure*}[htb!]
    \centering
    \includegraphics[width=0.8\linewidth]{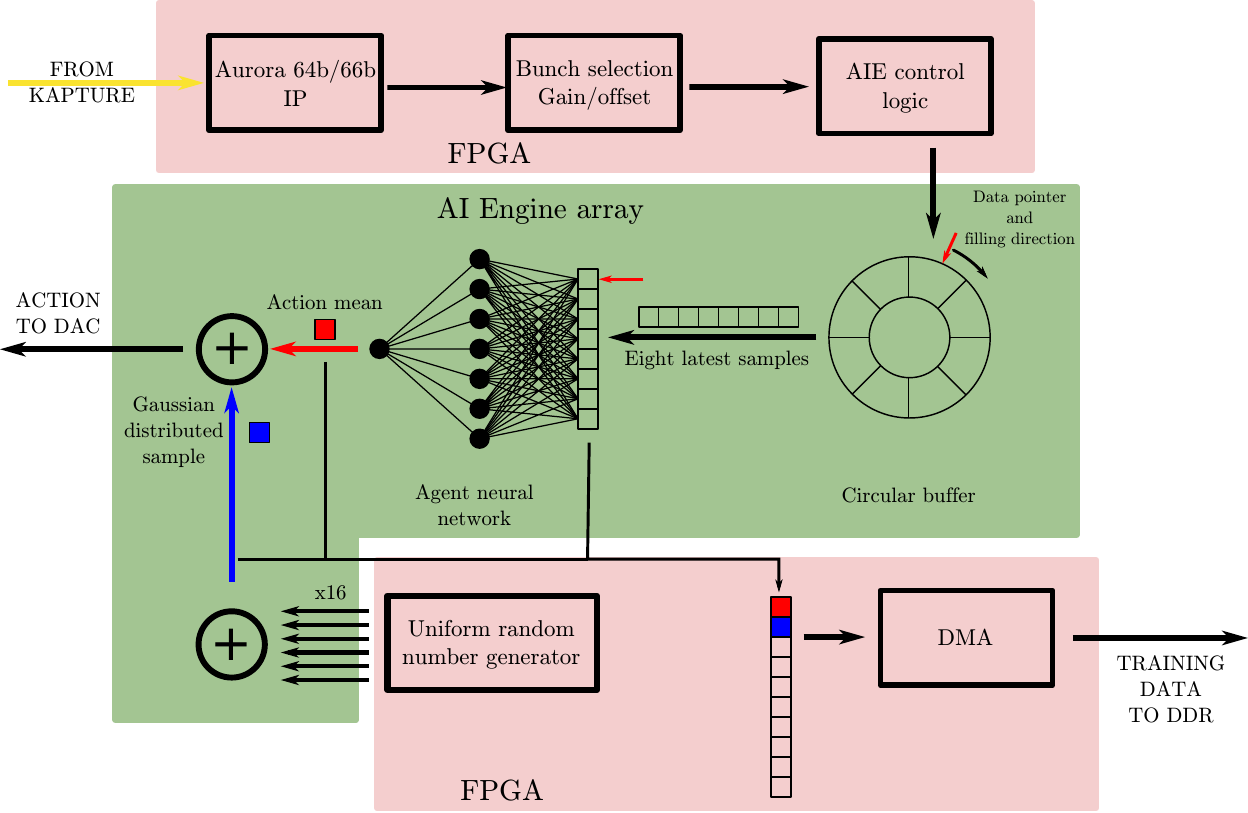}
    \caption{Schematic representation of the data path within the Versal VCK190 firmware. In the top part, the KAPTURE data stream coming from HighFlex 2 is decoded by an Aurora IP from the protocol used for fiber-optic communication. The bunch of interest for the control experiment is selected and gain/offset correction are applied. \ac{FPGA} logic takes care of forwarding data and gracefully stopping the \ac{AIE}. In the middle, a buffer stores the latest eight samples and feeds them to the \ac{NN}. The exploration noise is added and the action is then output to a DAC, for control of the kicker. In the bottom side, a \ac{FPGA} block takes the latest data from the inference and stores them in memory through a \ac{DMA} for later training.}
    \label{fig:versal_internal}
\end{figure*}

A schematic of the internal data processing within the actor and the KINGFISHER platform is shown in \cref{fig:versal_internal}. The first \ac{AIE} tile implements a circular buffer and streams the latest eight samples to the following kernel using the cascade stream interface. These eight samples represent the observation vector, the choice of which will be described more thoroughly in \cref{subsec:rl_problem_def}. The next kernel implements the linear layer of a \ac{NN} computing the values of the sixteen hidden neurons. A \ac{ReLU} activation function is applied to the outputs. Such an activation function was selected because of its simple implementation. This function can also be turned off while the system is still running in order to implement linear filters. The output of the network is then computed with a final linear layer and passed to the last output kernel. All these kernels keep forwarding the eight input values. The last kernel takes 16 values from uniformly-distributed random number data stream and sums them, producing an approximately Gaussian-distributed sample due to the central limit theorem. This is used to add a Gaussian noise with a standard deviation selectable at run-time to the output of the network. A final data vector containing the eight input values, the random value, and the output of the network previous to the noise addition is given to the experience accumulator logic. 

A computer in the control room then fetches the data and uses a modified version of the Stable-baselines3 library~\cite{stable-baselines3} \ac{PPO}~\cite{schulman2017proximal} implementation to train the policy using the hyper-parameters in \cref{tab:rl_parameters}. The new parameters are then loaded onto the \ac{AIE} kernel at run-time and new data for training is gathered. The whole inference loop has a latency of $\SI{2.8}{\micro\second}$. For this work, a number of $2048$ action steps was chosen. This number has been manually selected in such a way that the agent would not have enough time to cause beam loss and disrupt operation.

The critic network was chosen to be identical to the actor network. The remaining hyper-parameters are available in \cref{tab:rl_parameters}.

\begin{table}
\centering
\begin{tabular}{lr}
\toprule
Hyper-parameter         & Value \\
\midrule
Learning rate $\eta$   & 0.0012 \\
Discount rate $\gamma$ & 0.99   \\
Number of steps        & 2048   \\
Batch size             & 64     \\
Number of epochs       & 1      \\
\bottomrule
\end{tabular}
\caption{Hyper-parameters used for the \ac{PPO} algorithm.}
\label{tab:rl_parameters}
\end{table}

\section{Demonstration of real-time RL control at the Accelerator}

\subsection{Task description: Horizontal betatron oscillations at KARA}
The \acf{KARA} is a ramping electron storage ring with $\SI{0.5}{GeV}$ injection energy, with various operation modes between $\SI{0.5}{GeV}$ and $\SI{2.5}{GeV}$. Its lattice is based on four sectors of double-bend achromats, with a total circumference of $\SI{110}{m}$.

In synchrotrons, the horizontal beam dynamics is dominated at first order by the linear betatron motion. The bunch displacement from the reference orbit $x$ can be expressed with Hill's equation
\begin{equation}
      \frac{d^2 x}{ds^2} + K_x(s) x = 0  
\end{equation}
where $s$ is the length along the ring and $K_x$ are the periodic focusing functions.

The general solution to this equation is
\begin{equation}
\begin{split}
    x(s) & = x_0 \sqrt{\beta_x(s)} \cos[ \theta(s) + \theta_0] \\
    \theta(s) & = \int_0^s \frac{ds}{\beta_x(s)}
\end{split}
\end{equation}
where $\beta_x$ are the beta-functions and $(x_0, \theta_0)$ set the initial conditions. The number of full oscillations per revolution is defined as the horizontal tune $Q_x$. In injection mode, used for this experiment, $Q_x \approx 6.76$. An observer fixed in a given position of the accelerator will only observe the fractional part, corresponding to an oscillation frequency in the order of $\SI{700}{kHz}$, depending both on the operation mode, the beam current and more generally the state of the machine. This oscillation frequency sets the timescale at which the \ac{RL} agent must be able to act, in this case in the order of a few microseconds.

A stripline kicker, based on the design from~\cite{dehler2000current}, is capable of affecting the \acp{HBO} by applying a bunch-by-bunch and turn-by-turn horizontal force to the beam based on the signal provided into an analog input.

The injection in the \ac{KARA} storage ring makes use of three strong kicker and one septum magnet in order to properly merge the beam already in the machine with the one that is being injected. These kicker magnets can only activate at a rate of $\SI{1}{Hz}$, for a few revolutions, and can move the beam on a displaced orbit in the horizontal plane. When the kicker switches back off, the displaced bunches will start performing betatron oscillations around the reference orbit. The goal of the \ac{RL} agent is to damp this oscillation as quickly as possible. Notably, the strength of the injection kickers is orders of magnitude stronger than the stripline kickers used for feedback, thus an agent cannot damp an oscillation in a single kick, and turn-by-turn control is required.

A classical controller for this problem exists and is already available in commercial solutions. These \ac{BBB} feedback systems~\cite{bbbdimtel} are usually based on a \ac{FIR} filter that takes the input signal and applies a $\SI{\pi}{\radian}$ phase at the frequency of the instability. In this way, a linear kick is produced with an opposite sign compared to the displacement, in this way damping the oscillations. The output of this kind of filter can be computed as
\begin{equation}
    y(t)=\sum_{i=0}^{N} c_i x(t-i)
\end{equation}
where $c_i$ are the coefficients of the filter, and $N$ is its order. The tuning of the filter coefficients directly impacts the performance of the controller, as it defines its behavior with respect to external noise and the bandwidth over which a suitable phase offset is produced. So far this is usually hand-tuned. A review on the topic is provided in~\cite{yellowbookBBB}.

\subsection{Formulation as an RL task}
\label{subsec:rl_problem_def}
For the current problem of controlling the \ac{HBO}, the \ac{RL} environment was modeled as follows. Given the \ac{HBO} dynamics at a fixed position in the storage ring can be approximated by an harmonic oscillator, the position $x$ and its derivative $\dot{x}$ are sufficient to have full knowledge of the state of the system. In a discrete-time setting, the derivative can be computed from the time difference of two consecutive samples. 
\begin{equation}
    \dot{x}(t) = \frac{x(t) - x(t-1)}{\Delta t}
\end{equation}
This in turn means that the two latest position samples, $x(t)$ and $x(t-1)$, are also a full representation of the system's state. In practice, though, only having two values is subject to measurement noise. Thus, the observation vector was defined as the last eight positions $\bm{o}_t = (x(t), x(t-1), ..., x(t-7))^T$. The signal is sampled at the revolution frequency, i.~e.~ca.~$\SI{2.7}{MHz}$, thus eight samples span roughly two periods of the betatron oscillation.

The action is a force that is applied to the bunch through a stripline kicker. In the harmonic oscillator model, this corresponds to a driving force. Under this definition, the system is a \ac{MDP}.

Several different reward definitions were chosen, and the respective performance of the final agents are compared in \cref{subsec:ttrd}. All rewards studied in this paper penalize the agent when the $x$ position differs from zero, corresponding to the reference orbit.

\subsection{Simulation study}
In order to study the interaction of an agent with the \ac{HBO}, an environment based on the Gymnasium library~\cite{towers_gymnasium_2023} was developed. The dynamics was modeled as a damped harmonic oscillator with user selectable undamped angular frequency $\omega_0$ and damping ratio $\Gamma$. The environment stores the actions $a_i$ performed on the system and convolves this vector with the Green's function $B(t,t^\prime)$ of the damped harmonic oscillator

\begin{equation}
B(t,t^\prime)=\Theta(t-t^\prime) \frac{e^{-\Gamma (t-t^\prime)}}{\omega} \sin{\omega(t-t^\prime)},
\end{equation}
with $\omega=\sqrt{\omega^2_0-\Gamma^2}$. As such, the position $x(t)$ is computed as
\begin{equation}
x(t) = \sum_{i} B(t,(i+\Delta \tau)T_\text{rev}) \, a_i,
\end{equation}
where $\Theta(t)$ is the Heaviside step function and $T_\text{rev}$ is the revolution time of the accelerator.
An additional user selectable delay $\Delta \tau$ is added to the argument of the function to study the effect of latency. Gaussian noise is added to the samples, reflecting the behavior of real-life data. A kick of intensity one order of magnitude higher than what the agent can perform is applied at a random time to simulate the external kicker. 

In order to guide the selection of an \ac{RL} algorithm and agent structure, the environment was used to test the training performance with the algorithms available in the Stable-baselines3 library~\cite{stable-baselines3}. \ac{PPO} and the observation vector definition of \cref{subsec:rl_problem_def} were thus validated in simulation before testing the complete system on the accelerator. This is necessary in order to disentangle a hardware platform failure from an issue with the \ac{RL} problem formulation, in the case control could not achieved during the tests at KARA.

\subsection{Reinforcement Learning Control}
\begin{figure*}[htb!]
    \centering
    \includegraphics[width=\textwidth]{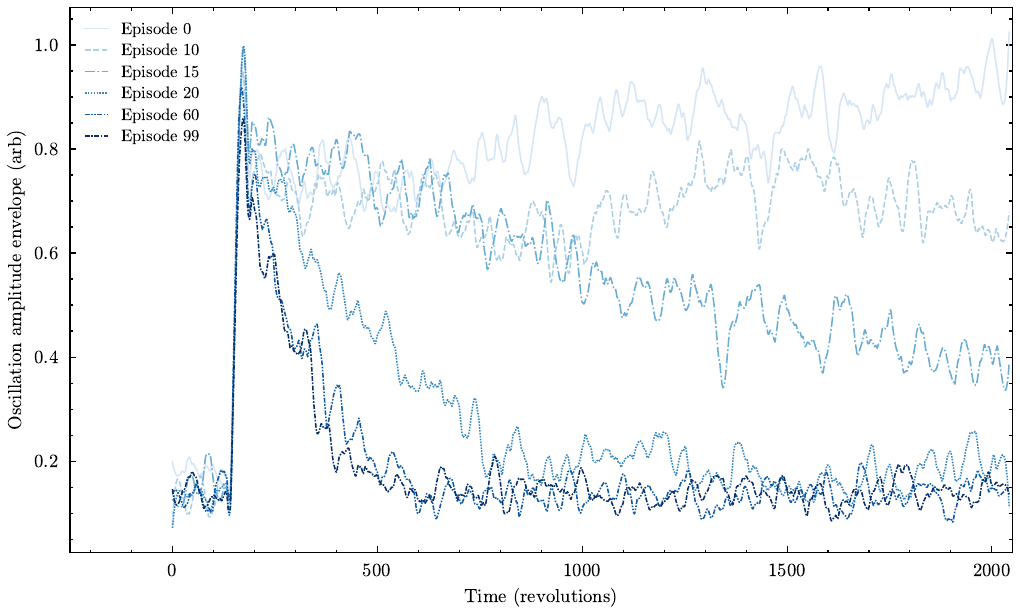}
    \caption{Smoothed envelope of oscillations measured by a \ac{BPM}. The sharp increase around $t=150$ revolutions is due to the injection kicker emulating an instantaneous external excitation. Notice how at step 0 the randomly selected agent is destabilizing the beam, leading to an increase of the oscillation amplitude. Moreover, the rate of damping increases with the number of training steps, i.e. with the agent experience.}
    \label{fig:envelope_improvement}
\end{figure*}

The system discussed in this study was allowed to interact with the accelerator for $2048$ revolutions (corresponding to $\SI{784}{\unit{\micro\second}}$). During this period an external kicker excited the oscillations as shown in \cref{fig:envelope_improvement}. After each of these episodes, a training step was performed, updating the coefficients of the \ac{NN}. The new set of weights and biases were uploaded to the agent and the operation was repeated.

\begin{figure}[htb!]
    \centering
    \includegraphics[width=\linewidth]{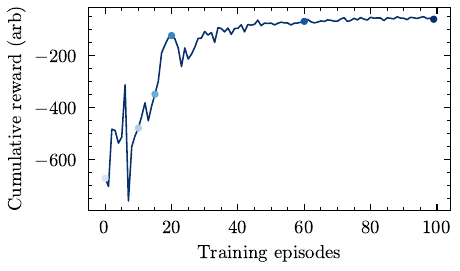}
    \caption{Cumulative reward obtained by an agent as a function of the number of training episodes. It can be seen how experience is gained (i.e. more episodes are used for training) and eventually plateaus. The colored points correspond to the episodes depicted in \cref{fig:envelope_improvement}.}
    \label{fig:reward_vs_step}
\end{figure}

In order to study the evolution of the oscillation amplitude, the amplitude of the oscillation was obtained as the absolute value of the Hilbert transform of the raw oscillation signal $x(t)$. As shown in \cref{fig:envelope_improvement} an increase in the damping rate of the oscillations is an indication that the agent in question is achieving control of the environment. Moreover, it is possible to examine the trend of the cumulative reward obtained during each training episode as shown in \cref{fig:reward_vs_step}. A clear increase in the obtained cumulative reward is visible as more episodes are used for training. This clearly shows that the agent improves with experience, as it is expected.

Several different training configurations were tested, each one with a different reward definition and the number of neurons in the hidden layer of the actor. This led to agents with different performances. An example of training configuration is L2, 12 N, meaning the L2 norm defined in \cref{tab:reward_defs} is used, together with an actor having 12 neurons in the hidden layer.

\subsection{Training Time Reward Definition}
\label{subsec:ttrd}

\begin{figure}[htb!]
    \centering
    \includegraphics[width=\linewidth]{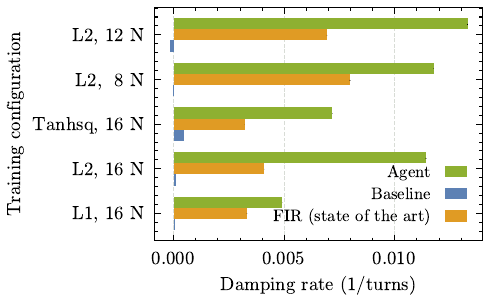}
    \caption{Damping rates of an instantaneous external excitation that is achieved for different kinds of training parameters. The $y$-axis denotes the reward function used for the specific training according to \cref{tab:reward_defs} and the number of neurons in the hidden layer. The trained and untrained (baseline) \ac{RL} agents are compared against an \ac{FIR} controller. Note that the negative baseline values are caused by the agents with random coefficients actually exciting the instability. All agents outperform the state-of-the-art \ac{FIR} controller showing higher damping rates.}
    \label{fig:agent_comparison}
\end{figure}

\Cref{fig:agent_comparison} compares the performance of different reward functions employed during training. To do so, the oscillation amplitude was fitted with an exponential function
\begin{equation}
f(t; A,\lambda) = A e^{-t \lambda}
\end{equation}
and the damping rate $\lambda$ was employed as a reward independent metric. Provided $x$ is the position obtained from the \ac{BPM}, the reward functions definitions are listed in \cref{tab:reward_defs}. 

\begin{table}[!ht]
\centering
\begin{tabular}{lr}
\toprule
Reward name & Definition \\
\midrule
L1     & $-|x|$                   \\
L2     & $-x^2$                   \\
Tanhsq & $-\tanh\left(x^2\right)$ \\
\bottomrule
\end{tabular}
\caption{Definition of the different reward functions used experimentally, where $x$ denotes the transverse horizontal position of the beam read by the \ac{BPM}.}
\label{tab:reward_defs}
\end{table}

The agents trained with all of these three reward choices reached a final performance better than the FIR controller and the baseline with the untrained agents.

\subsection{Online Network Structure Reconfiguration}
In order to further increase the level of flexibility of the system, the possibility of dynamically modifying the \ac{NN} structure without the need of re-implementing or re-packaging the firmware was implemented. This was achieved by embedding a smaller network into one with a greater number of neurons and layers by appropriately switching off different weights. Additionally, maintaining the number of computations constant allows to have identical latency between different training trials, thus removing this variability when comparing different agents.

Agents with several different layer sizes have been trained and their performance is shown in \cref{fig:agent_comparison}. The performance of all agents increased with training, outperforming the traditional \ac{FIR} controller. The best performing agent, with 12 neurons in the hidden layer and trained with an L2-norm reward (in short notation: L2, 12 N; cf.~also \cref{tab:reward_defs}), is used throughout the rest of this work for comparison with classical control techniques.

\subsection{Training stability and robustness}
\begin{figure}[!htb]
    \centering
    \includegraphics[width=\linewidth]{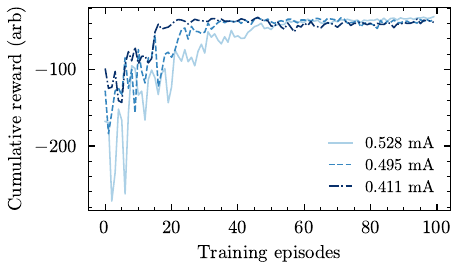}
    \caption{Comparison of different training episodes with different beam current conditions, where the final agents achieve almost identical performances.}
    \label{fig:training_stability}
\end{figure}

The training procedure was repeated several times, with the same setting but a different beam current, to study the stability of the agents produced. One would expect an effect for two reasons. First of all the \ac{BPM} signal is not normalized, so the amplitude will vary with current. Second of all the \ac{HBO} tune is current dependent~\cite{Blomley2021_1000137621}. Despite a $20\%$ reduction in beam current due to the natural decay of the beam, all resulting agents achieve a very similar final reward, as shown in \cref{fig:training_stability}. This shows the robustness of the \ac{RL} agent against variations of current. 

\begin{figure}[htb!]
    \centering
    \includegraphics[width=\linewidth]{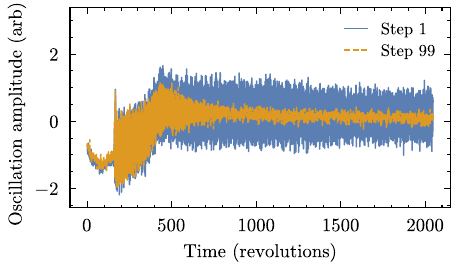}
    \caption{Evolution of the horizontal position of the beam after an instantaneous external excitation at time $t\approx\SI{175}{revolutions}$, influenced by an RL agent before and after training. The septum magnet was active, inducing a baseline shift superimposed on the usual exponential decay. Nonetheless, the trained agent (orange) is capable of damping the oscillation compared to the untrained one (blue).}
    \label{fig:septum_train}
\end{figure}

One of the main components of the \ac{KARA} storage ring injection line is the injection septum magnet. Its impulse activation is necessary to guarantee the injection of the electron bunch coming from the booster into the main ring. The leaking magnetic field, though, also affects the beam that is already in the storage ring. This effect is visible in \cref{fig:septum_train}, which corresponds to a shift in the position of the beam. Such an effect was not present in the simulation, but it was nonetheless possible to train an agent capable of correctly handling this new phenomenon. This is an example of the versatility and adaptability of \ac{RL} algorithms, that are sometimes able to autonomously learn from situations they are not originally designed for.

\subsection{Improvement during cumulative reward plateau}
\begin{figure}[htb!]
    \centering
    \includegraphics[width=\linewidth]{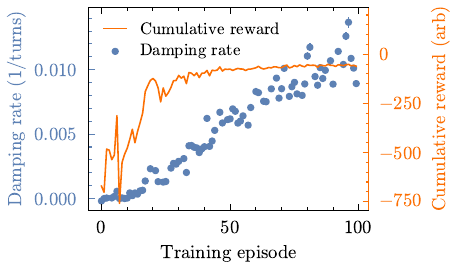}
    \caption{Damping rate (blue) measured with a different BPM system as a function of the training step, compared with the reward function (orange). Notice the absence of a plateau in the damping rate curve.}
    \label{fig:nonsaturation}
\end{figure}

As can be seen in \cref{fig:training_stability}, the cumulative reward reaches a plateau around step number 50. Nonetheless, if one studies the trend of the damping rate measured at a \ac{BPM} in a different part of the ring, it is still possible to observe an increase of the damping rate even around step 100 as shown in \cref{fig:nonsaturation}. This is due to the fact that noise in the input data adds an offset to the cumulative reward that hides small improvements. Such a phenomenon needs to be considered in future experiments as it could potentially hinder further improvement of the agent.

\section{Discussion}
We trained several working agents and their performance can be compared and evaluated. %

As can be seen in \cref{fig:agent_comparison}, the performance of the \ac{FIR} controller is not constant. This behavior is mainly due to variations in the beam current that, due to its linear nature, affects the action signal amplitude. This is not the case for the \ac{RL} agent, as it is able to automatically adapt to the variation in beam current. Thus, the trained agent outperformed the \ac{FIR} controller in all of our evaluations. Additionally, the untrained \ac{RL} agent is shown as the baseline, with clearly inferior performances when compared to the trained agent and the \ac{FIR} controller.

\begin{figure}[!thb]
    \centering
    \includegraphics[width=\linewidth]{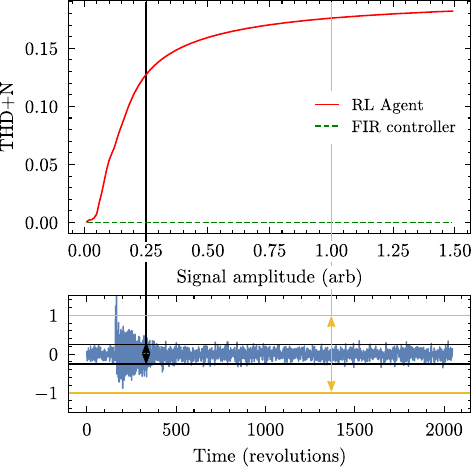}
	\caption{Total harmonic distortion plus noise (THD+N) as a function of the input signal amplitude for an (L2-norm, 12 neurons) agent.  Note that the \ac{FIR} filter with negligible noise will have a (THD+N) close to 0. In the bottom plot, a training signal is shown, allowing to determine which part of the training episode employs a higher non-linear behavior. The noise level is indicated with the black lines, while an higher amplitude is shown with the yellow lines. Signals above the noise level tend to produce more non-linear actions.}
    \label{fig:thd_plot}
\end{figure}

Compared to the \ac{FIR} controllers conventionally used, \ac{NN} agents are capable of exhibiting non-linear output response. This allows the implementation of more complex policies. Particularly shallow \acp{NN} with \ac{ReLU} activation functions have been shown to behave linearly in some cases. When pure sinusoidal input is fed into a black-box, the non-linear behavior produces harmonics of the fundamental sinusoidal input. The \ac{THD+N}, is defined as
\begin{equation}
    \text{THD+N} = \frac{\sqrt{A^2_N + \sum_{i=2}^\infty A^2_{i}}}{A_1},
\end{equation}
where $A_i$ is the amplitude of the $i$-th harmonic, where $i=1$ is the fundamental, and $A_N$ is the noise amplitude. This metric expresses the amount of non-linear components in the output of a given device. For a linear controller, like a \ac{FIR} filter, in the case where noise is negligible, $\text{THD+N}$ is approximately zero. The \ac{THD+N} was computed for different amplitudes of a sinusoidal input at the betatron oscillation frequency.  The behavior for the L2, 12 N agent is shown in \cref{fig:thd_plot}. The amount of non-linear harmonic content is consistent, with a steep increase at a level compatible with the noise floor of the signal provided to the agent. This might indicate that the agent is learning to apply more complex actions in the case of high-amplitude, and thus highly penalized, observations.

The trade-off between training through interaction with a simulated or real-world environment is an important aspect of the application of \ac{RL} to large-scale facilities. A simulation-driven approach is sometimes necessary to ensure safety, both of the facility and its personnel, while in other cases, it is dictated by the time necessary to obtain the training dataset. One advantage of real-world training is that trained agents are directly transferable into operation. This is not the case for agents trained on simulation, as their transfer to the real world is potentially hindered by non-modeled phenomena and, more in general, differences between the training and real-world environments. This work presents the opportunity to compare these two approaches. To do so, the time necessary to train an agent through interaction with the accelerator and simulation is reported in \cref{tab:training_time}. It is important to consider that the online case comprised waiting for a $\SI{1}{Hz}$ trigger, controlling the kickers. These numbers comprise the overhead of transferring training data, the time necessary for \ac{NN} back-propagation and the interaction time with the environment. Given both approaches used the same model, the back-propagation time can be assumed to be equal. The interaction time, on the other hand, is $\SI{17.6}{s}$ for the simulation, while the trial on the accelerator takes $\SI{0.076}{s}$ or $\SI{76}{ms}$. It is relevant to notice how the simulation being employed, and discussed in greater detail in the method section, is lightweight while still performing two orders of magnitudes slower than gathering data on the machine. As such, approximately $50\%$ of the real-world training time is consumed by data-access overhead. This could be reduced in future implementations of the system with several approaches: using higher speed network links paired with lower overhead network protocols, or by accelerating the training directly on the FPGA and \ac{AIE} array sharing memory with the experience accumulator hardware. In conclusion, for systems with dynamics that is computationally intensive to simulate, the techniques described in this article will greatly improve the time necessary for training. An environment-driven training procedure, i.e. training directly on the real-world task, becomes thus not only possible but also more flexible than a simulation study as the total deployment time would be reduced.

\begin{table}
\centering
\begin{tabular}{lcr}
\toprule
Training platform & Interaction time & Training time \\
\midrule
Simulation CPU    & $\SI{17.6}{s}$   & $\SI{137}{s}$ \\
Simulation GPU    & $\SI{17.6}{s}$   & $\SI{227}{s}$ \\
Online CPU        & $\SI{0.076}{s}$  & $\SI{260}{s}$ \\
\bottomrule
\end{tabular}
\caption{Comparison of the time necessary to perform 100 training steps for different training platforms.}
\label{tab:training_time}
\end{table}

In certain scenarios characterized by rapid dynamics and computationally intensive simulations demanding high-performance computing clusters, it may be conceivable that training directly on the accelerator consumes less energy than utilizing simulations. Such a possibility could significantly influence the sustainability of \ac{ML} methodologies.

\section{Conclusions}
In this study, we introduced the experience accumulator architecture and the training time reward definition, marking a significant step in implementing real-world on-the-edge environment learning for \ac{RL}-based controllers. Particularly in scenarios where simulation costs are prohibitive and data generation rates are high, this methodology emerges as a promising solution, enabling the deployment of \ac{RL} controllers. Our application of this approach to real-time control of particle accelerator dynamics yielded inference latency of a few microseconds.

Utilizing advanced heterogeneous computing platforms such as KINGFISHER, our presented implementation facilitates the deployment of plug-and-play \ac{RL} systems operating at microsecond latency scales. This opens the door to intelligent control of ultra-fast non-linear dynamics~\cite{scomparin:ipac2024-tups61} in systems such as particle accelerators and fusion experiments. The efficacy of these capabilities was demonstrated through the control of the \ac{HBO} at the accelerator test facility KARA, resulting in a functional controller comparable and outperforming state-of-the-art FIR controllers used during standard operation phases of synchrotron light sources.

\section{Acknowledgements}
The authors thank Lars Eisenblätter, Jennifer Derschang and Alexander Bacher for their assistance with the electronics used in this work. Luca Scomparin and Chenran Xu acknowledge the support by the DFG-funded Doctoral School “Karlsruhe School of Elementary and Astroparticle Physics: Science and Technology”.
Andrea Santamaria Garcia acknowledges funding by the BMBF ErUM-Pro project TiMo (FKZ 05K19VKC).

\section{Code and Data Availability}
All data and code used for this work are provided by the authors upon reasonable request.

\bibliography{main}

\end{document}